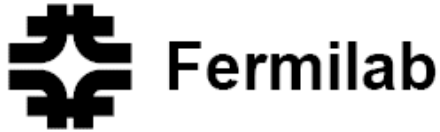

# DETECTOR BACKGROUND AT MUON COLLIDERS[*][†]

N.V. Mokhov[#], S.I. Striganov

Fermi National Accelerator Laboratory, Batavia, IL 60510, USA

## Abstract

Physics goals of a Muon Collider (MC) can only be reached with appropriate design of the ring, interaction region (IR), high-field superconducting magnets, machine-detector interface (MDI) and detector. Results of the most recent realistic simulation studies are presented for a 1.5-TeV MC. It is shown that appropriately designed IR and MDI with sophisticated shielding in the detector have a potential to substantially suppress the background rates in the MC detector. The main characteristics of backgrounds are studied.

[*]Work supported by Fermi Research Alliance, LLC under contract No. DE-AC02-07CH11359 with the U.S. Department of Energy.
[†]Presented at TIPP2011 Conference, Chicago, June 2011, to be published in *Physics Procedia Journal*.
[#]Corresponding author. E-mail: mokhov@fnal.gov

# 1. Introduction

Muon Collider (MC) detector performance is strongly dependent on the background particle rates in various sub-detectors. Deleterious effects of the background and radiation environment produced by muon decays are one of the fundamental issues in the feasibility study of the MC ring, Interaction Region (IR), Machine-Detector Interface (MDI) and detector. Although incoherent $e^+e^-$ pair production at the Interaction Point (IP) and beam loss on limiting apertures can result in noticeable background levels, muon decays have been identified as the major source of detector backgrounds at a MC [1-3]. The decay length for a 0.75-TeV muon is $\lambda_D = 4.7 \times 10^6$ m. With $2 \times 10^{12}$ muons in a bunch, one has $4.28 \times 10^5$ decays per meter of the lattice in a single pass, and $1.28 \times 10^{10}$ decays per meter per second for two beams.

Electrons from muon decays have mean energy of approximately 1/3 of that of the muons. At 0.75 TeV, these ~250-GeV electrons, generated at the above rate, travel to the inside of the ring magnets, and radiate a lot of energetic synchrotron photons tangent to the electron trajectory. Electromagnetic showers induced by these electrons and photons in the collider components generate intense fluxes of muons, hadrons and daughter electrons and photons, which create high background and radiation levels both in a detector and in the storage ring at the rate of 0.5-1 kW/m. This is to be compared to a good practice number of a few W/m at superconducting (SC) hadron colliders.

Detector performance is affected by backgrounds in three ways: detector component radiation aging and damage, difficulties with reconstruction of objects (e.g., tracks) not related to products of $\mu^+\mu^-$ collisions, and deterioration of detector resolution (e.g., jet energy resolution due to extra energy from background hits).

# 2. Earlier Findings and New Studies

In very early MC studies, it was found that [1-4]:
- Photon, electron and neutron fluxes and energy deposition in detector components are well beyond technological capabilities if one applies no measures to bring these levels down.
- Tungsten nozzles, starting a few centimeters from IP with ±20-deg outer angle, are the most effective way (~1/500) of background suppression.
- The nozzles can also fully confine incoherent pairs if $B_{detector} > 3$ T.
- High-field SC dipoles implemented in the final focus region, interlaced with quadrupoles and tungsten masks, provide further reduction of backgrounds.
- With such an IR design, the major source of backgrounds in a MC detector is muon decays in the region reduced to about ±25 m from the IP.
- Time gates promise substantial mitigation of background problem in a MC detector (quantified below in section 4.5).

These findings have recently been confirmed and the MC background problems further attacked in coherent studies by collider lattice and magnet designers, particle production and transport experts and detector groups [5, 6]. A consistent design now exists for a compact $0.75 \times 0.75$ TeV $\mu^+\mu^-$ collider ring, IR, MDI and chromatic correction section with large momentum acceptance and dynamic aperture, all based on high-field $Nb_3Sn$ SC magnets adequately protected against dynamic heat loads.

## 3. MARS Modeling of Backgrounds

Energy deposition and detector backgrounds are simulated with the MARS15 code [7]. All the related details of geometry, materials distributions and magnetic fields for lattice elements and tunnel in the ±200-m region from IP, detector components [8], experimental hall and MDI are implemented in the model (Fig. 1). To protect the SC magnets and detector, 10 and 20-cm tungsten masks with 5 $\sigma_{x,y}$ elliptic openings in the IR magnet interconnect regions and sophisticated tungsten cones inside the detector [1, 2] were implemented (yellow in Fig. 1) into the model and carefully optimized. The 0.75-TeV muon beam is assumed to be aborted after 1000 turns. The minimal cut-off energies in this study range from 0.001 eV for neutrons to 1 MeV for muons and charged hadrons. The cut-off energy in the tunnel concrete walls and soil outside is position-dependent and can be as high as a few GeV at 50-100 m from the IP compared to the minimal value in the vicinity of the detector.

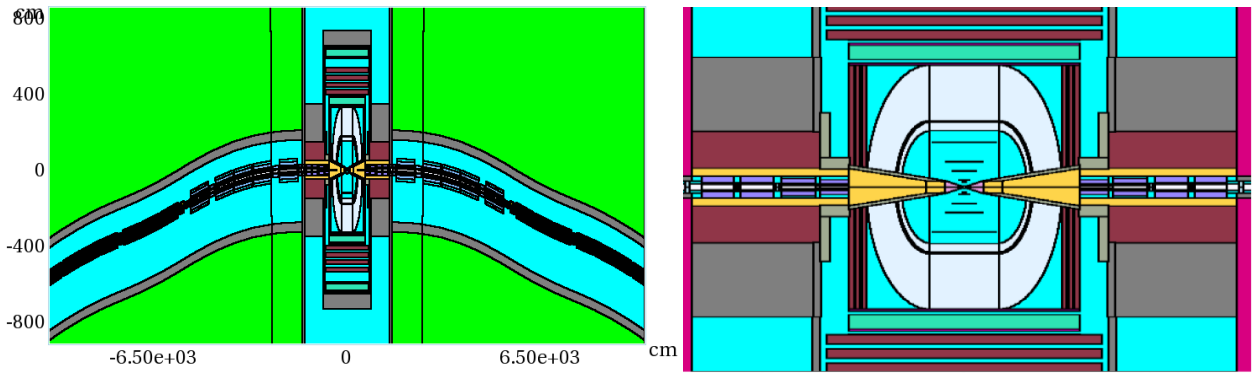

Fig. 1. Plan views of MARS model for IR (left, axes in cm, |z|<10000 cm) and detector with MDI (right, |x|<500 cm, |z|<1500 cm).

Fig. 2 (left) shows muon flux isocontours in the MC IR. These muons – with energies of tens to hundreds of GeV - illuminate the entire detector. They are produced in the Bethe-Heitler process by energetic photons from electromagnetic showers generated by decay electrons in the lattice components. The neutron isofluences inside the detector are shown in Fig. 2 (right). The maximum neutron fluence and absorbed doses in the innermost layer of the silicon tracker for a one-year operation are at a 10% level of that expected in the LHC detectors at the nominal luminosity.

The dipoles close to IP and tungsten masks in each interconnect region help reduce background particle fluxes in the detector by a substantial factor. The tungsten nozzles, assisted by the detector solenoid field, trap most of the decay electrons created close to IP as well as most of incoherent $e^+e^-$ pairs generated in the IP. Their outer angle in the region closest to IP (6 to 100 cm) is the most critical parameter to optimize: the larger this angle the better background suppression, but the impact on the detector performance, especially in the forward region, becomes higher. The total numbers of photons and electrons entering the detector per bunch crossing are $1.5 \times 10^{11}$ and $1.4 \times 10^{9}$, respectively, for the minimal studied outer angle of the nozzle of 0.6 degrees, and reduced by three orders of magnitude for the MDI with the angle of 10 degrees. Results in the rest of this paper are for the 10-degree configuration. Note that the inner opening shape and dimensions are carefully optimized in the nozzle region of 6 < |z| < 600 cm, and 5-cm thick borated polyethylene cladding is used at 100 < z < 600 cm where the outer angle is reduced to 5 degrees.

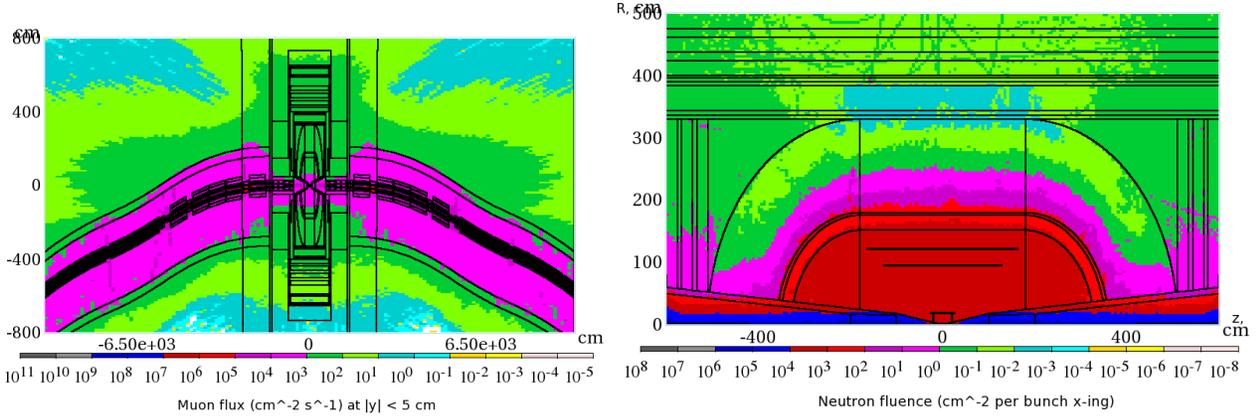

Fig. 2. Muon isofluxes in IR (left) and neutron isofluences in the detector (right).

In the MARS15 runs, a source term for detector simulations is calculated for all particles entering the detector through the MDI surface. This surface is defined around IP (r=2.2 cm z=±13 cm), on the outer surface of the nozzle up to z=±600 cm, and on the inside of the detector at r=655 cm z=±750 cm. Corresponding high-statistics files have been generated for two 0.75-TeV muon beams, with minimal variation of particle weights, and with full information on the particle origin. The main characteristics of this source term are described in the next section (for cut-off energies indicated in Table 1).

## 4. Main Characteristics of Backgrounds

### 4.1. Origin in lattice

As found in earlier studies and confirmed with the current IR and MDI designs, the origin of all particles (except muons) entering the detector is the straight section of about ±25 m near the IP. The combined effect of angular divergence of secondary particles, strong magnetic field of dipoles in IR and tight tungsten masks in interconnect regions is that there is practically no contribution to non-muon detector backgrounds from distances z > ±25 m (see Fig. 3, left). Excellent performance of the optimized nozzles and MDI shielding along with confinement of decay electrons in the aperture (forcing them to hit the nozzle on the opposite side of IP) result in the longitudinal distributions of particle origins shown in Fig. 3 (left), with a broad maximum from 6 m to 17 m (IP side of the first dipole).

On the contrary, Bethe-Heitler muons hitting the detector are created in the lattice as far as 200 m from IP, with 90% of them produced at ±100 m around IP (Fig. 3, right). As for all other particles, the fine structure of these distributions is related to the lattice details, with pronounced peaks always connected to the high-field SC dipole locations.

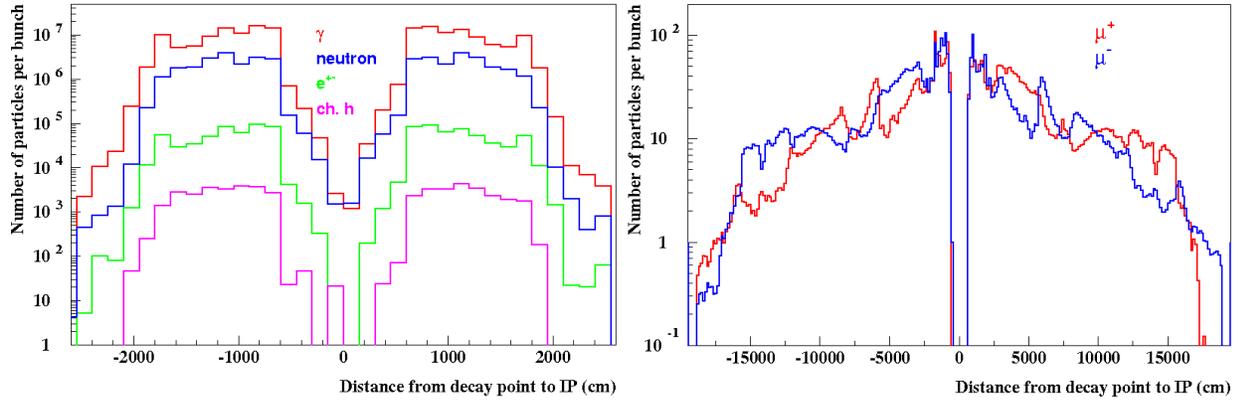

Fig. 3. Numbers of background particles (per bunch crossing) entering the detector as a function of distance along the IR lattice to their production point: Bethe-Heitler muons (right) and other particles (left).

*4.2. Spatial distributions at interface surface*

Fig. 4 shows longitudinal point-of-entry distributions of backgrounds for a positive muon beam. Most particles enter the detector through the nozzle outer surface. The maximum yield of photons and electrons is very close to the IP where the shielding is minimal. Neutron and charged hadron yields peak at $z = \pm 1$ m. Muons enter the detector through the $z = \pm 750$ cm plane (70%) and the nozzle (30%).

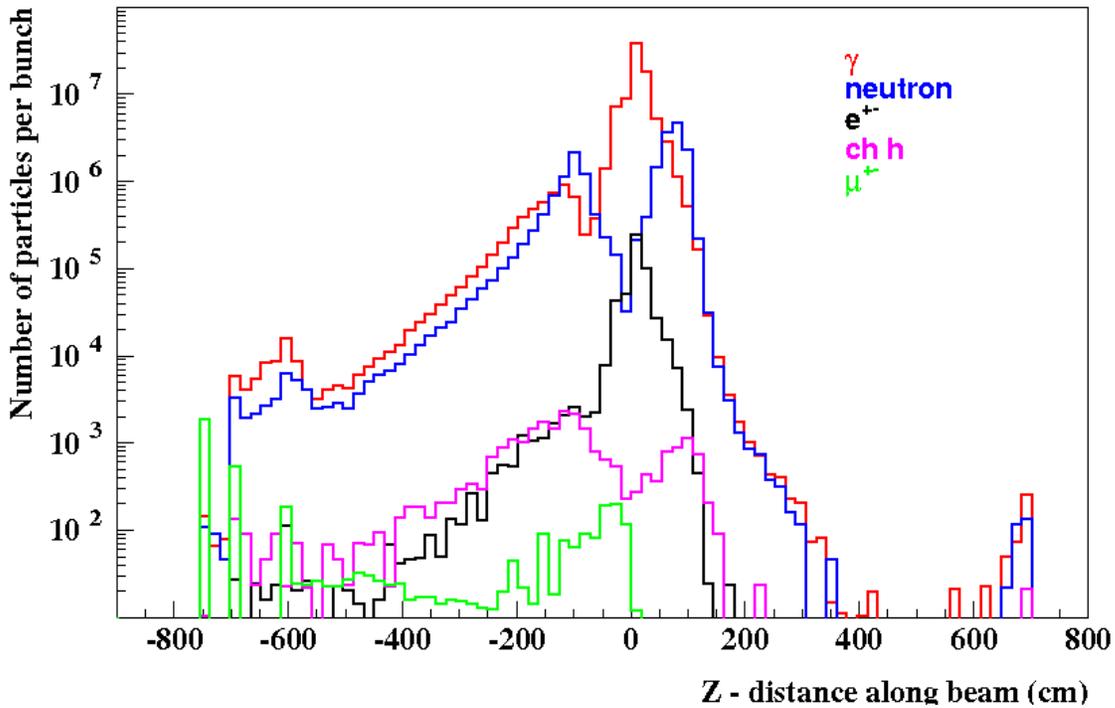

Fig. 4. Longitudinal distributions of entry points of background particles (per bunch crossing) for $\mu^+$ beam moving from the left.

The fact that most particles (except muons) are produced close to the IP and not affected by the strong dipole magnetic field results in the azimuthal symmetry of the source term with the corresponding distributions at MDI being flat. Angular distributions of these particles on the interface surface also have the azimuthal symmetry. On the contrary, Bethe-Heitler muons are strongly affected by the magnetic fields of the dipoles on their long way to IP. As a result, azimuthal distributions of these muons have a strong asymmetry as shown in Fig 5 for the $\mu^+$ beam. Secondary positive muons are deflected by the IR element magnetic fields to the same side as a positive muon bunch (negative horizontal direction). Secondary negative muons are deflected to the opposite side. Note, that many muons hit the detector at large radii.

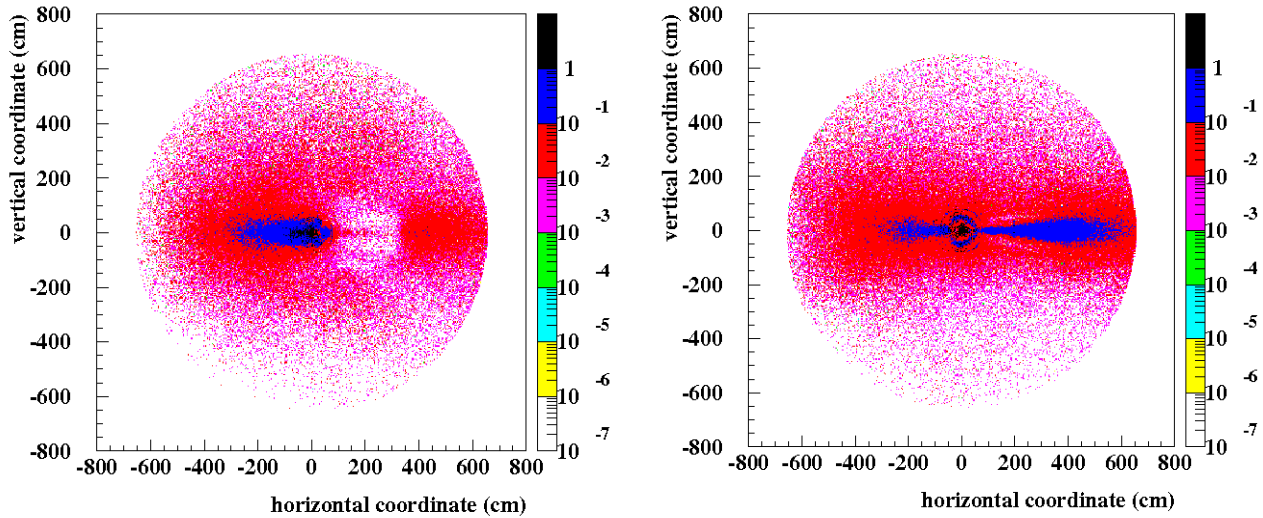

Fig. 5. $\mu^+$ (left) and $\mu^-$ (right) distributions at the detector entrance for the $\mu^+$ beam.

*4.3. Particle and energy flows into detector*

The total numbers of particles (with cut-off energies $E_{th}$ indicated) entering the detector through the MDI surface are given in Table 1 along with particle mean momenta and energy flow. One sees that soft photons and neutrons are the major components. They are followed by electrons and positrons. Mean momenta of background particles are rather low except for charged hadrons (~0.5 GeV/c) and Bethe-Heitler muons (22 GeV/c). About 540 TeV of energy is brought to the detector by background particles per bunch crossing. Photons, neutrons and muons contribute about one third each to the energy flow, with two other components being small.

Table 1. Mean numbers <n>, momenta <p> and total kinetic energy flow <EF> of background particles per bunch crossing

| Particle ($E_{th}$, MeV) | <n> | <p>, MeV/c | <EF>, TeV |
|---|---|---|---|
| Photon (0.2) | $1.8 \times 10^8$ | 0.91 | 164 |
| Neutron (0.1) | $4.1 \times 10^7$ | 45 | 172 |
| Electron/positron (0.2) | $1.0 \times 10^6$ | 6.0 | 5.8 |
| Charged hadron (1) | $4.8 \times 10^4$ | 513 | 12 |
| Muon (1) | $8.0 \times 10^3$ | 23030 | 184 |

*4.4. Momentum spectra*

As one can see from Fig. 6, the momentum spread of particles entering the detector through the MDI surface is quite broad. With the kinetic cut-off energies indicated above used, photons and electrons with their ~MeV/c mean momenta, have always $p < 0.2$ GeV/c. Hadron momentum reaches ~3 GeV/c, with very different mean values for neutrons and charged hadrons (see Table 1). Bethe-Heitler muons, illuminating the detector, do have the highest momenta of up to 200 GeV/c.

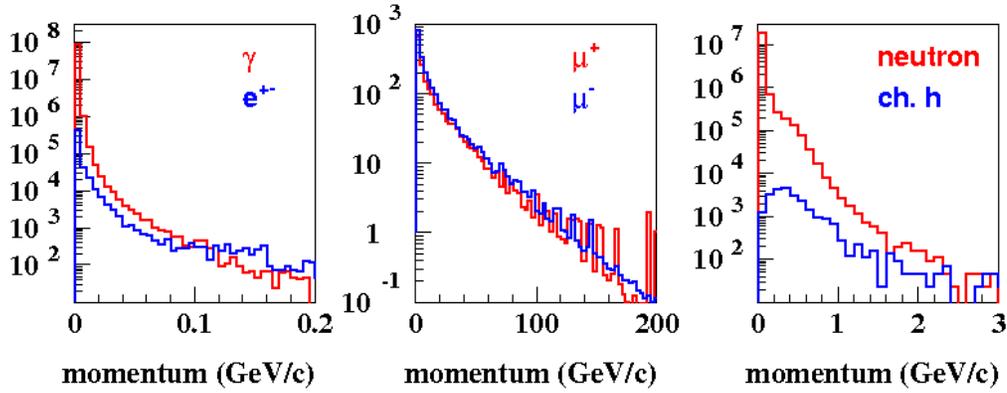

Fig 6. Momentum spectra at MDI for photons and electrons (left), muons (middle) and hadrons (right).

*4.5. Time distributions*

The time of flight (TOF) of background particles at the MDI surface has a significant spread with respect to the bunch crossing as shown in Fig. 7. Two regions are clearly seen in the TOF distributions. The first one at TOF < 40 ns is related to the direct contributions from particles generated by muon beam decays in the ±17 m region not shielded by the strong magnetic field of the first dipole (see Fig. 3). The long tails for photons, electrons/positrons and neutrons are due to their bouncing and multiple interactions in MDI components at low energies. The long tail for energetic Bethe-Heitler muons is associated with their production at large distances from IP (Fig. 3, right). These properties of the TOF distribution of the source term at MDI suggest that one can use timing in the detector to reduce the number of the readout background hits. As shown in Ref. [9], the background neutron hit rate registered in vertex and tracking silicon detectors can be reduced by a factor of several hundred by using the 7-ns time window.

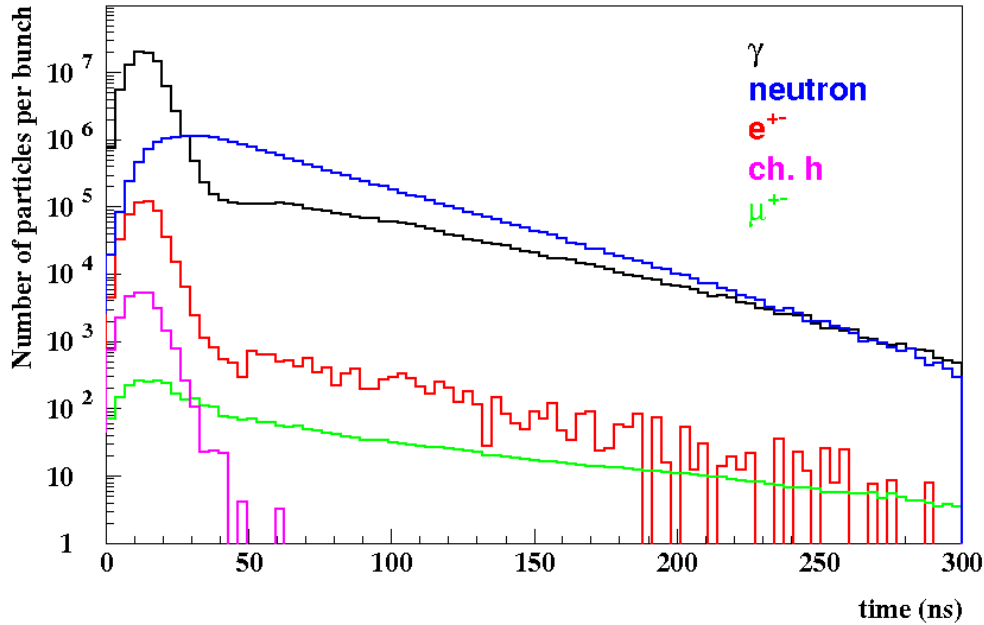

Fig. 7. Time of flight distributions of background particles at the detector entrance with respect to bunch crossing.

## 5. Conclusions

The recent developments in the design of the 1.5-TeV center-of-mass Muon Collider ring and interaction region (based on 8-10 T $Nb_3Sn$ superconducting magnets) along with substantial efforts in Monte-Carlo code developments and optimization studies of the machine-detector interface and appropriate detector technologies assure one that the severe background environment at such a challenging machine can be reduced to tolerable levels. The main characteristics of particle backgrounds entering the collider detector are studied in great details. A good understanding of backround properties suggests the ways for suppression of the detector response to the non-IP related hits.

## Acknowledgements

We would like to thank Y.I. Alexahin, V.Y. Alexakhin, V. Di Benedetto, R. Carrigan, C. Gatto, S. Geer, V.V. Kashikhin, R. Lipton, A. Mazzacane, N. Terentiev and A. Zlobin for collaboration and fruitful discussions.